\documentclass[sigconf, nonacm=True]{acmart}
\AtBeginDocument{%
  }

\setcopyright{acmlicensed}
\copyrightyear{2018}
\acmYear{2018}
\acmDOI{XXXXXXX.XXXXXXX}
\begin{document}

\title{The Evolving Landscape of Interactive Surface Sensing Technologies}


\author{David Wang}
\affiliation{%
  \institution{University of Michigan}
   \city{Ann Arbor}
   \state{Michigan}
   \country{USA}}
\email{davwan@umich.edu}

\author{Wilson Chen}
\affiliation{%
  \institution{University of Michigan}
   \city{Ann Arbor}
   \state{Michigan}
   \country{USA}}
\email{wilchen@umich.edu}

\author{Tianju Wang}
\affiliation{%
  \institution{University of Michigan}
  \city{Ann Arbor}
  \state{Michigan}
  \country{USA}}
\email{tianjuw@umich.edu}

\author{Jiale Zhang}
\affiliation{%
  \institution{University of Michigan}
  \city{Ann Arbor}
  \state{Michigan}
  \country{USA}}
\email{jiale@umich.edu}


\begin{abstract}
Interactive surfaces have evolved from capacitive touch and IR based systems into a diverse ecosystem of sensing technologies that support rich and expressive human computer interaction. This survey traces that progression, beginning with infrared vision based approaches, such as FTIR and diffuse illumination, and the rise of capacitive touch as the dominant technology in modern devices, to focusing on contemporary modalities including vision and acoustic sensing. New technologies under development are also discussed, including mmWave radar, and vibration based techniques. Each sensing technique is examined in terms of its operating principles, resolution, scalability, and applications, along with discussions of multimodal integration. By comparing tradeoffs between sensing modalities, the survey highlights the technical and design factors that shape interactive surface performance and user experience. The review concludes by identifying persistent challenges, including sensing accuracy, power constraints, and privacy concerns, and outlines how emerging sensing modalities can enable future interactive environments to be ubiquitous and intelligent.
\end{abstract}




\maketitle


\section{Introduction}

Over the past two decades, interactive surface technologies have undergone a systematic evolution from early optical imaging approaches to capacitive touch sensing and, more recently, to multimodal perception. Distinct sensing modalities have followed diverging technical trajectories, each balancing trade-offs in spatial resolution, scalability, material compatibility, and environmental robustness. These developments have collectively driven a shift from two dimensional touch input toward richer, spatially expressive forms of interaction. This review follows that trajectory, synthesizing the major sensing paradigms that underpin contemporary interactive surfaces and examining the challenges and opportunities that shape their continued advancement.

Early interactive surfaces relied heavily on infrared and vision-based sensing, with techniques such as Frustrated Total Internal Reflection (FTIR) \cite{Han2005} and Diffuse Illumination (DI) \cite{Wilson2004, Wilson2005} offering low cost, scalable solutions for large format multi-touch systems that supported early tabletop, tangible, and public information interfaces. However, these optical approaches were highly sensitive to ambient lighting, surface contamination, and imaging noise, and their precision and form factor limitations constrained broader deployment \cite{Sheridan2010, Go2012}. As mobile computing advanced, capacitive touch sensing emerged as the dominant consumer technology: mutual capacitance architectures \cite{Lee2014} resolved the multi-touch ambiguity of self-capacitance and enabled thin, high precision, low latency touch panels now standard across smartphones, tablets, and wearables. Yet capacitive sensing remains restricted by material dependencies, environmental variability, and limited expressiveness for force, shear, and richer tactile modalities \cite{Harrison2012Touche, Ismail2021CapacitiveReview, Nature2023SoftSkin}.

Over the past decade, rapid advances in visual algorithms, acoustic hardware, and AI models have significantly expanded the capabilities of interactive surfaces. Computer vision has progressed from projector–camera augmented tabletops \cite{10.1145/1268517.1268539} to spatial interaction in XR systems \cite{10.1145/3698140, Rosales_2019}, where SLAM and depth sensing enable real-time understanding of both the surrounding environment and user pose. Acoustic sensing has advanced in parallel both in speech driven interaction through modern ASR systems \cite{baevski2020wav2vec, nayeem2025automatic} and in touch inference based on impact and vibration propagation \cite{goel2014surfacelink, 1037150, lopes2011augmenting} allowing reliable input even under poor lighting or challenging material conditions.

At the same time, emerging modalities such as mmWave and vibration sensing show considerable promise. mmWave radar can recognize fine-grained gestures, trajectories, and even multi-user movement with millimeter level precision \cite{lien2016soli, pegoraro2021real}, while providing a more privacy preserving alternative to camera based sensing \cite{zhang2023survey}. Vibration based approaches exploit mechanical wave propagation through structures to enable touch localization, force estimation, and activity recognition on ordinary furniture, walls, and wearable devices \cite{10.1145/3544548.3580991, 7944790, tanaka2015wearable}.

Despite these advances, deploying sensing systems in real environments remains challenging. Key issues include generalization across users and materials, power constraints, privacy and data security, the stability of non-contact interaction, and the complexity introduced by multimodal fusion. Achieving robust, scalable, and privacy-aware interaction will be essential for broader adoption.

Against this backdrop, this review examines the major sensing modalities that support interactive surfaces, including infrared and optical methods, capacitive touch, vision-based and acoustic sensing, and emerging mmWave and vibration techniques. We discuss their principles, capabilities, limitations, and applications, and highlight cross-cutting challenges and future research directions to inform the design of next-generation, intelligent, and ubiquitous interactive surfaces.

\section{Background}
Several sensing modalities have long served as the foundation for interactive surfaces, with infrared-based optical systems and capacitive touch technologies standing out as the most influential. These mature and widely deployed approaches form the core of this section, providing essential historical context for understanding how modern surface interaction first took shape.

\subsection{Infrared} 
Optical and infrared imaging approaches played a pivotal role in the early development of interactive surface technologies. 
Among the various optical sensing approaches, FTIR (Frustrated Total Internal Reflection) and DI (Diffuse Illumination) emerged as the two most influential techniques. Together, these two techniques formed the foundation of vision-based multi-touch sensing and played a central role in establishing the conceptual and technical foundations of large scale interactive surfaces.

\subsubsection{Frustrated Total Internal Reflection}

\paragraph{Principles and Applications}
FTIR (Frustrated Total Internal Reflection)is a touch sensing mechanism grounded in optical reflection behavior. Its operation relies on total internal reflection (TIR), a phenomenon that occurs when light propagates within a high refractive index medium and encounters a boundary with a lower refractive index at an angle exceeding the critical angle. Under this condition, light is completely reflected back into the medium. When a finger or another object touches the surface, the local interface is transformed from acrylic–air to acrylic–skin, changing the refractive index contrast and disrupting the TIR condition. This disruption allows light to escape and scatter at the contact point. An infrared camera positioned beneath or beside the panel captures these bright scattered regions, enabling the system to recover touch locations from the imagery \cite{Han2005}.

\begin{figure}
    \centering
    \includegraphics[width=1\linewidth]{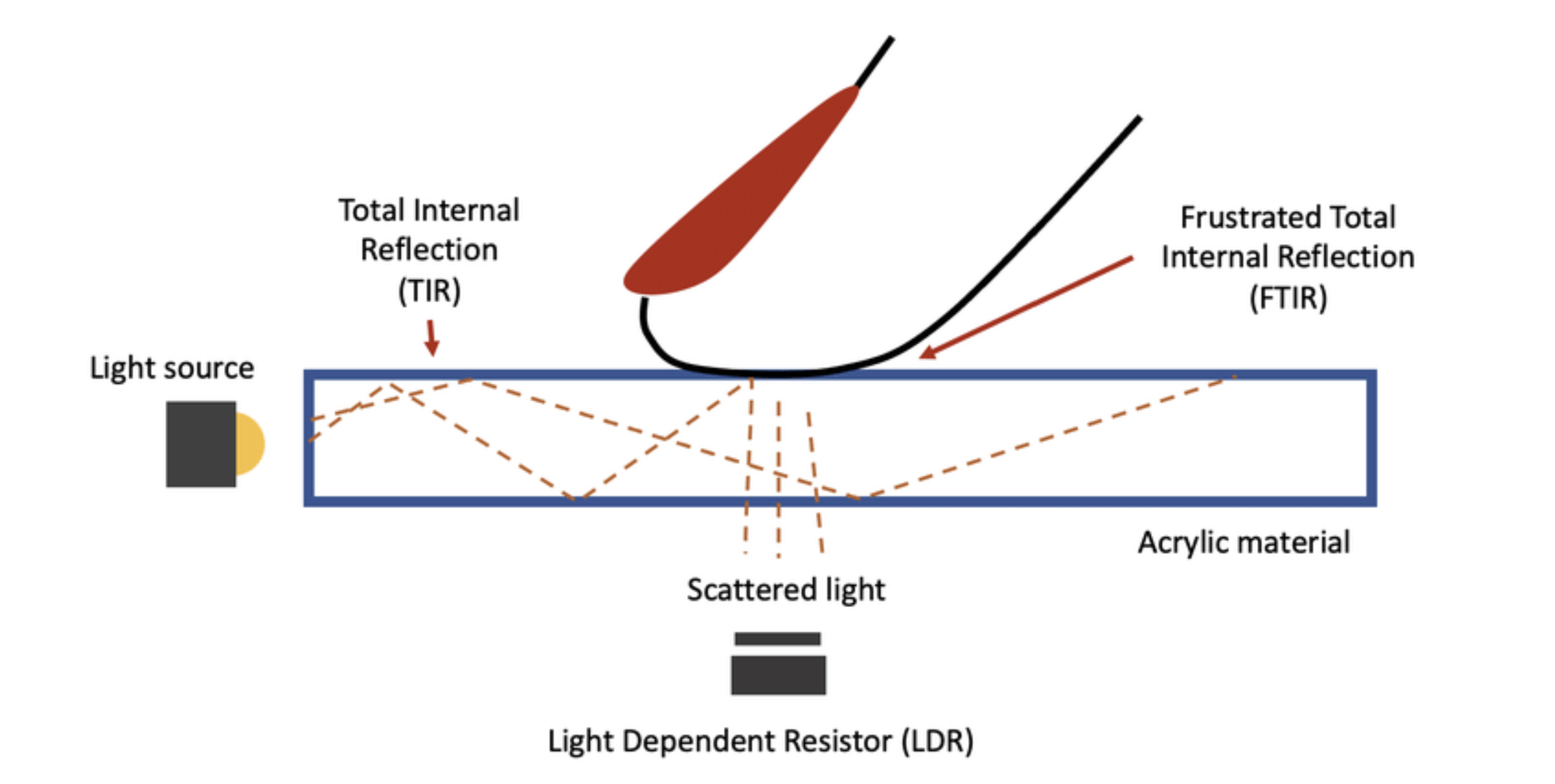}
    \caption{An FTIR touch sensing system \cite{everitt2022supporting}.}
    \label{fig:ftirtouch}
\end{figure}

FTIR emerged as a central foundation for early interactive surface and multi-touch research due to its flexibility, low cost, and ease of scaling. Han’s UIST 2005 system \cite{Han2005} demonstrated the promise of this approach through smooth multi-finger interaction and highly responsive visual feedback. The work is widely regarded as an early catalyst for the multi-touch movement and inspired both commercial platforms such as Microsoft PixelSense and sustained exploration of low cost optical touch systems within communities such as the NUI Group. Leveraging the advantages of optical propagation across large substrates, FTIR also supported early interactive walls and immersive display systems. Examples include HoloWall \cite{Matsushita1997}, TouchLight \cite{Wilson2004}, and PlayAnywhere \cite{Wilson2005}, which highlighted the feasibility and scalability of optical interaction in public information spaces and portable projection environments.

In collaborative settings, FTIR’s inherent ability to support simultaneous multi-point input enabled shared workspaces, map-based interaction, and educational applications. DiamondTouch \cite{Dietz2001} was a notable milestone in this area, establishing mechanisms for user differentiation and demonstrating how multi-user interaction could be reliably supported on a single tabletop system. Owing to its low construction barrier, reliance on common components, and system openness, FTIR also became a preferred platform for research prototypes and experimental interface design. This accessibility helped enable influential interdisciplinary projects such as the Reactable musical interface \cite{Jorda2007} and the TUIO framework \cite{Kaltenbrunner2007}, further strengthening FTIR’s role as a key building block in the early development of interactive surface technologies.

\paragraph{Limitations and Challenges}
Although FTIR played a formative role in early interactive surface systems, its reliance on disrupting total internal reflection limits reliability and applicability. Stable signals occur only under sufficient pressure or contact area, making light touches and stylus like inputs difficult to detect, a weakness consistently noted in vision-based touch studies \cite{Han2005,Jacucci2010}. The need for a thick acrylic light guide further restricts FTIR to bulky, rigid form factors, preventing use in thin or mobile devices.

Environmental sensitivity compounds these issues: dust, moisture, and strong ambient illumination introduce light leakage or reduce infrared contrast, undermining segmentation and robustness in public or outdoor settings \cite{Wilson2005,Kaltenbrunner2007}. Spatial accuracy is also limited because touch points are inferred from variable light blobs influenced by posture, skin properties, and material imperfections, hindering fine-grained or pen-based input~\cite{Dietz2001,Jorda2007}.

Despite these constraints, FTIR remains valuable for large, low-cost, multi-user installations, continuing to influence research and artistic deployments through its optical simplicity and accessibility.

\subsubsection{Diffuse Illumination}
\paragraph{Principles and Applications}
Diffuse Illumination (DI) represents another major class of infrared imaging–based touch sensing techniques. Its central principle is to infer touch events by observing how finger contact alters the distribution of diffuse infrared light across the interaction surface. In contrast to FTIR, which depends on disruptions of total internal reflection, DI focuses on changes in global illumination uniformity or local reflectance properties. Touch points are detected by identifying local decreases or increases in brightness, which are then segmented using well-established computer vision methods such as background subtraction, thresholding, and connected component analysis \cite{Stauffer1999,Zivkovic2004}. This approach enables DI to function effectively on large interactive surfaces without requiring a light guiding substrate.

Within the DI family, Rear DI is one of the earliest and most widely adopted variants. In this configuration, infrared light is projected upward from beneath the surface to create a uniform diffuse field. When a finger approaches or touches the surface, it occludes a portion of the upward directed illumination, producing a shadow or darkened region in the camera’s view. Touch detection is thus achieved by locating these occlusion induced attenuations, forming what is often described as an “inverse FTIR” mechanism. Wilson’s TouchLight system leveraged volumetric shadows to capture hand gestures and touch input \cite{Wilson2004}, and his subsequent PlayAnywhere project demonstrated the portability of Rear DI for table, wall, and floor based interaction scenarios \cite{Wilson2005}. Rear DI also played a central role in Underkoffler’s Illuminating Light systems, which showed how diffuse illumination could scale to room sized environments and support full body interaction.
\begin{figure}
    \centering
    \includegraphics[width=1\linewidth]{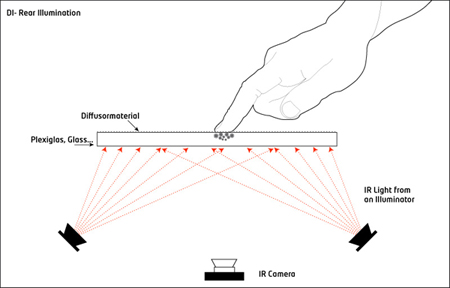}
    \caption{An example of Rear DI for touch sensing \cite{hanmulti}.}
    \label{fig:ditouch}
\end{figure}

Front DI constitutes the second major DI variant and differs from Rear DI in its reliance on reflectance rather than occlusion. Infrared light is projected from above or at an oblique angle onto the interaction surface. When a finger touches the surface, the reflective and scattering characteristics of human skin cause the contact region to appear as a bright spot in the camera image. This “reflectance enhancement” mechanism resembles the operation of optical mouse sensors, but applied to an entire surface rather than a localized patch. Matsushita and Rekimoto’s HoloWall system captured reflected IR features to support combined proximity and touch input \cite{Matsushita1997}. Front DI has also been widely used in systems involving object recognition; for example, the Reactable employed camera-based detection of both reflected touch highlights and fiducial markers to unify touch and tangible interaction \cite{Jorda2007}. These systems collectively illustrate how DI enables versatile touch and gesture sensing without imposing strict material constraints on the interaction surface.

\paragraph{Limitations and Challenges}

Diffuse Illumination (DI), although attractive for large and material flexible interactive surfaces, faces several constraints inherent to its optical sensing model. Its most fundamental limitation is strong sensitivity to ambient illumination: shadow cues in Rear DI quickly degrade under strong or multidirectional lighting, while reflectance cues in Front DI are easily overwhelmed by environmental infrared, making controlled lighting almost a requirement \cite{Wilson2004,Irri2014,Gershon1986}. Rear DI also suffers from intrinsic shadow ambiguity hands, arms, and nearby objects frequently generate false activations, and even background modeling techniques such as mixture of Gaussians struggle under fluctuating lighting \cite{Stauffer1999,Go2012}. Front DI encounters complementary issues, as dust, oils, moisture, and stray reflective elements produce bright artifacts that closely resemble true touches; field studies show such reflective noise is far more severe in real deployments than in controlled environments \cite{Sheridan2010,Jorda2007}, despite mitigation efforts such as external light evasion techniques \cite{Evasion2014}.

In addition, DI inherits the spatial-resolution limits of camera-based sensing: touch points are estimated from brightness centroids that vary with posture, reflectance, and illumination, preventing the high precision achievable with capacitive sensing. Consequently, DI is ill-suited for fine-grained input or uncontrolled lighting conditions, though it remains valuable for large, low-cost, and installation-based interactive systems.

\subsection{Capacitive} 

\subsubsection{Evolution and Applications}

Capacitive touch sensing is \\grounded in the principle of capacitive coupling: when a human finger, as a conductive body, approaches or contacts a sensing electrode, it perturbs the local electric field and induces measurable changes in capacitance. By monitoring these changes, the system infers the presence and location of touch inputs. In its early years, capacitive sensing was primarily adopted in industrial control panels and public use terminals such as elevator buttons and ticketing kiosks where its durability and reliability offered advantages over mechanical actuators \cite{Nam2021}. With the advent of high sensitivity capacitive front end circuits and low power ASICs, the technology entered consumer devices in the 1990s. Notebook trackpads, which employed self-capacitance detection by measuring the capacitance of each node relative to ground, quickly replaced mechanical pointing devices. However, self-capacitance suffered from well-known “ghosting” issues under multi-touch conditions, limiting its ability to support more sophisticated gesture inputs.

In the early 21st century, mutual capacitance matrix sensing emerged as the dominant architecture. By arranging transmit (TX) and receive (RX) electrodes in an intersecting grid and detecting coupling changes at each intersection, mutual capacitance allows inherently robust multi-touch and high-resolution input. Lee et al. (2014) demonstrated that electrode pattern optimization plays a decisive role in determining the sensitivity and signal characteristics of projected capacitive panels \cite{Lee2014}. Ko et al. (2021) further proposed a mutual capacitance readout IC synchronized with high refresh rate AMOLED drivers to mitigate noise and bandwidth constraints \cite{Ko2021}. In parallel, advances in transparent conductive materials such as metal mesh and silver nanowires enabled thinner, more integrated electrode structures compatible with direct display integration. These developments broadened capacitive sensing applications: capacitive touchscreens became the primary interaction modality for smartphones and tablets, supporting widely adopted gestures such as pinch, rotate, and swipe; in notebook computers, touchpads enabled multi-finger scrolling and gesture interaction; in large format whiteboards, wearables, and flexible devices, capacitive sensing expanded through larger electrode pitches or deformable electrode substrates.

\begin{figure}
    \centering
    \includegraphics[width=1\linewidth]{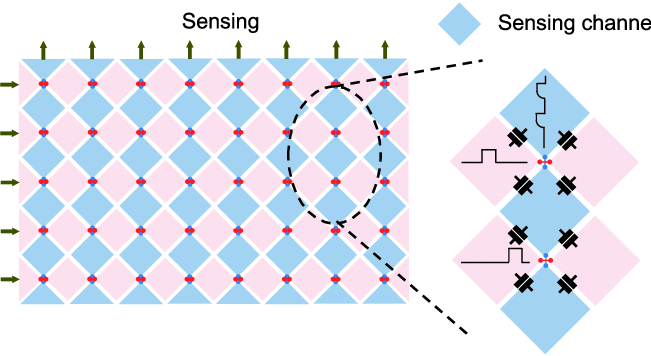}
    \caption{An implementation of a mutual capacitance matrix \cite{huang2018frequency}.}
    \label{fig:captouch}
\end{figure}

From a technical perspective, capacitive touch systems benefit from short electrical signal paths and rapid sampling, enabling high precision and low latency. Because they do not rely on optical imaging, they remain robust under varying lighting conditions. Furthermore, the ability to integrate electrodes directly into cover glass or display films contributes to thin form factor, highly integrated device designs. Mutual capacitance architectures inherently support multi-touch gestures, and the technology has matured into a cost efficient solution with a well-established global supply chain. These combined advantages have led capacitive touch technology to replace optical approaches such as FTIR and DI in most mainstream products, establishing it as the dominant sensing paradigm in modern mobile and interactive devices.

\subsubsection{Limitations and Challenges}

Capacitive coupling systems for skin device interaction still encounter substantial limitations that hinder their deployment in practical settings. Traditional surface capacitance architectures inherently constrain expressive bandwidth, making it difficult to reliably distinguish higher dimensional inputs such as force variations, sliding versus tapping, or complex gestures without introducing denser electrode matrices and more sophisticated signal processing pipelines, which considerably increase system complexity and engineering cost \cite{Harrison2012Touche,Ismail2021CapacitiveReview}. Their sensing performance is further affected by strong dependence on skin biophysical states and environmental conditions: moisture, sebum, contact area, finger orientation, and surface contaminants all perturb the electric field distribution and lead to instability in tactile feedback and touch recognition. For wearable and skin conformal devices, long term drift arises from variations in stratum corneum hydration, surface morphology, and continuous mechanical deformation an issue repeatedly emphasized in the wearable sensing literature \cite{WearableReview2020}. Moreover, structural trade-offs among sensitivity, spatial resolution, and multimodal tactile capability remain challenging. While flexible materials and microstructured dielectrics can enhance sensitivity and mechanical compliance, they also introduce higher fabrication complexity, reduced durability, and limited environmental tolerance \cite{Wiley2022CapSensors}. Achieving richer tactile modalities such as simultaneous pressure, shear, and deformation sensing remains difficult for conventional capacitive layouts, and studies on soft capacitive skin highlight persistent limitations including weak shear discrimination and poor conformity of rigid electrode structures to curved surfaces \cite{Nature2023SoftSkin}. Addressing these constraints generally demands hydrogels, stretchable conductors, multilayer electrode arrays, and algorithmic compensation, all of which significantly elevate fabrication and system integration barriers and continue to impede scalable real world adoption.

\section{Current Technologies}
Some of the most prevalent advances in sensing have come in visual and acoustic modalities. Many papers have taken steps to Mark Weiser's vision of ubiquitous computing, where computing devices are seamlessly integrated into everyday life \cite{10.1145/329124.329126}. The concept of a natural user interface (NUI) formed to describe such user interfaces that are intuitive and invisible. Modern embedded systems have seen a greater focus in hands-free technology that utilizes visual and acoustic sensing. Advances in hardware have allowed devices to compute and store large volumes of data at high speed, enabling researchers to take advantage of  more complex methods.

\subsection{Vision} 
Computer vision is a form of artificial intelligence that allows computers to interpret visual data from the world to identify objects and make decisions. Large volumes of data are collected to create a training dataset for a machine learning model. Convolutional neural networks have been the primary model for image processing, while recurrent neural networks are used for sequential data such as video frames. Advances in AI have given rise to the use of vision transformer models that split an input image into patches and use a self-attention mechanism to interpret the image, much like a language transformer using tokens. In past work, projection-based systems with cameras have been used to provide interactive visible surfaces. Current research trends towards extended reality (XR) devices that track the user's body orientation to interact with virtual and physical elements. Deployment in commercial interactive systems have become more common with services like Amazon Go, which uses sensor fusion between camera arrays and weight sensors to determine which products the customer takes.

\subsubsection{Physical Interactive Surfaces}

\begin{figure}
    \centering
    \includegraphics[width=1\linewidth]{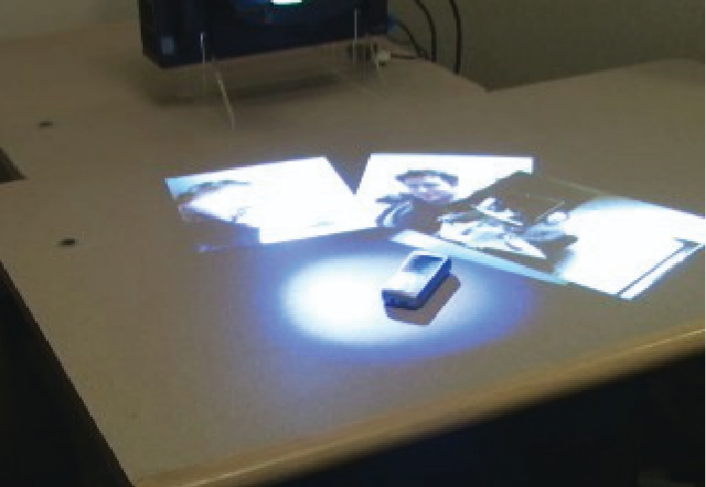}
    \caption{BlueTable's projection system connects to the phone on the table and displays interactive photos. \cite{10.1145/1268517.1268539}}
    \label{fig:bluetable}
\end{figure}

\paragraph{Projection}
Projection-based vision systems on tabletops have been used to enable interactive interfaces with multi-touch capabilities. These systems prioritize convenience through ease of use and portability. Early vision-based systems used a simple setup where a sheet of infrared light was projected onto a surface and detected fingers that interrupted the light. Researchers at Canesta utilized this to create an infrared projected keyboard \cite{10.1145/792704.792732}. Dynamic keystroke detection algorithms were used to detect typing. Wilson's previously mentioned PlayAnywhere work built upon this by incorporating a projector and a camera with an infrared-pass filter to detect finger positions \cite{Wilson2005}. An optical flow algorithm is also used to detect movements for natural interactions with project items. The follow-up paper on BlueTable extends this by enabling cellphones on the tabletop to connect to the system through Bluetooth (Figure \ref{fig:bluetable}). The connection is initiated by a visual handshake system detected by the camera, allowing users to interact with photos and perform file transfers with motion gestures \cite{10.1145/1268517.1268539}. Although projection-based systems offered portability and easy setup, the effectiveness of image processing depends on the projection surface, limiting where it can be deployed.

\paragraph{Retrofitted Displays}
Another approach involves retrofitting existing displays with cameras to make them into interactive surfaces \cite{5958988}. These improve upon projection-based systems since it does not require any special hardware or infrared technology. An interactive application runs on the computer while a webcam faces the display. Transformations are used to ensure that the received image of the display is a rectangle and background subtraction is done during a calibration stage. A histogram based skin color detection approach is used to identify the user's hand regions. Other objects placed on the display can become interactive objects through contour finding algorithms. While this may not be portable due to needing a display and computer, it serves as an inexpensive method to convert existing displays into an interactive surface.

\subsubsection{Virtual Interactive Surfaces}
\begin{figure}
    \centering
    \includegraphics[width=1\linewidth]{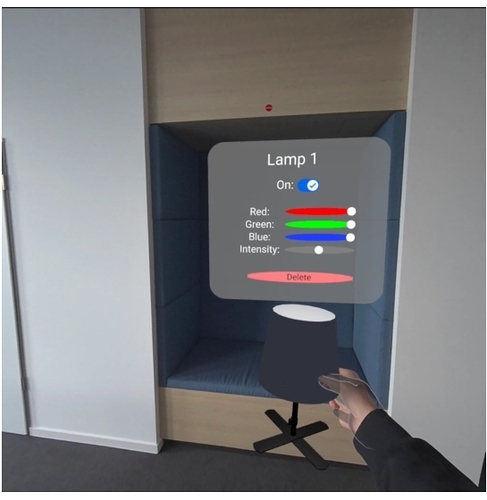}
    \caption{An example of smart home planning and automation through mixed reality and digital twin techniques\cite{10.1145/3733155.3733198}.}
    \label{fig:smarthome}
\end{figure}

Virtual surfaces have been another method to get around the static nature of traditional displays. In more recent years, research has focused on ways to connect the digital and physical world together through extended reality, which encompasses virtual, augmented, and mixed realities. Virtual reality (VR) immerses the user into a simulated 3D world with pose tracking and interaction. A fully virtual environment allows application developers to create interactive surfaces that work mid-air by tracking the user's hand and eye movement. Augmented reality (AR) provides a digital overlay onto the real world. Mixed reality combines the two and allows users to interact with both virtual and physical objects. Mixed reality (MR) allows users to have real-time interactions between digital and physical elements through a combination of computer vision and sensor suites that track user movement (Figure \ref{fig:smarthome}). Dynamic helmet-mounted vision systems are used to realize this, allowing computer vision to infer the interactive surfaces themselves.

\paragraph{SLAM Algorithms}
Simultaneous Localization and Mapping (SLAM) is the fundamental method for keeping track of one's position and orientation in XR technology. SLAM algorithms use sensor fusion from stereo cameras and depth sensors to map out the environment and determine where the user is in it. This process is known as perception, and this is used to understand the surfaces that are around the user. The act of determining where the user is located is localization, which is done using image processing on frames captured by the cameras. Research on SLAM for VR applications have been limited since raw sensor data from commercial head-mounted displays (HMDs) are not directly accessible due to their proprietary nature. SLAM simulation testing has also struggled due to high computational overhead when dealing with image features. However, recent work by Pinheiro et al. addresses this by performing SLAM in runtime virtual environments while utilizing mesh geometry for efficiency rather than iamge-based features \cite{desousa2025mesh2slamvrfastgeometrybased}.

\paragraph{Surface Interaction}
Once the environment has been interpreted, digital user interfaces can be placed around the environment to interact with through controllers or by hand tracking. With depth perception, this allows floating UI windows to be added around the physical environment for users to press. To improve the responsiveness, acoustic and haptic feedback (when using a controller or glove system) is given to simulate a physical response. A more recent concept is the use of spatial anchors to lock virtual surfaces to physical locations \cite{9417665}. This allows the surfaces to persist when out of view and can be helpful in multi-user situations. Peripheral information can also be diffused into a physical environment by mapping traditional 2D application windows onto objects \cite{10.1145/3626472}.

\paragraph{Applications}
As XR hardware continues to mature, a large emphasis has been put on practical applications. It can be used to assist in parametric architectural design or interact with virtual flight simulator interfaces \cite{10.1145/3698140, 10.1145/3626481}. 3D objects can also be modeled by using ribbon-like strokes from VR brushes \cite{Rosales_2019}. XR interactive surfaces often combine visual, acoustic, and haptic feedback to improve user response. These transform interactive surfaces into interactive spaces, where the user's complete surrounding becomes part of their interface.

\begin{figure}
    \centering
    \includegraphics[width=1\linewidth]{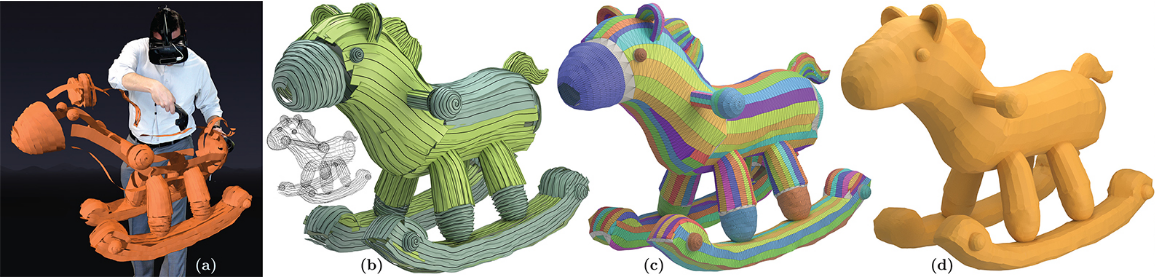}
    \caption{VR brush strokes are processed to determine intended surfaces and constructed into a 3D model \cite{Rosales_2019}.}
    \label{fig:surfacebrush}
\end{figure}


\subsection{Acoustic}   
Acoustic centric systems make use of sounds produced by humans or objects as a source of data for interaction and sensing. Acoustic approaches in interactive surfaces has two primary modes: voice based control/interaction interfaces, and hardware based tactile interfaces which make use of audio data for sensing. Voice-based audio commands are typically powered by natural language processing (NLP), and include systems such as Amazon Alexa, Siri, and newer LLMs, and serve as one of the primary methods for interaction with modern devices. Acoustic sensing hardware makes use of sounds from user actions to identify gestures and movements, which can be used to control a smart space.
\subsubsection{Voice Control Interfaces}
\paragraph{Natural Language Recognition}
Interactive surfaces and environments often make use of Automatic Speech Recognition (ASR) for human computer interaction. ASR has traditionally been implemented with Fast Fourier Transforms (FFTs) and other spectral analysis techniques for feature extraction, which can then be used to train machine learning models. After raw audio data is parsed into written language, semantics based decision trees can then be used to turn recognized sentences into actions the device can perform. Newer ASR systems make use of neural networks, with audio encoders used to process audio data directly into features \cite{nayeem2025automatic}, without the need for spectral analysis. While traditional ASR systems rely heavily on labeled speech to recognize words, advanced techniques such as wav2vec can learn internal speech representations directly from raw audio data using self supervised learning while capturing temporal knowledge through the use of transformers, improving speech understanding and reducing training data requirements \cite{baevski2020wav2vec}. Transformer based audio to speech recognition is used by multimodal LLM systems such as Google's Gemini and OpenAI's GPT-4o \cite{team2023gemini}, which have been used as aids for manipulating extended reality interactive environments \cite{chen2025analyzing}.
\begin{figure}
    \centering
    \includegraphics[width=1\linewidth]{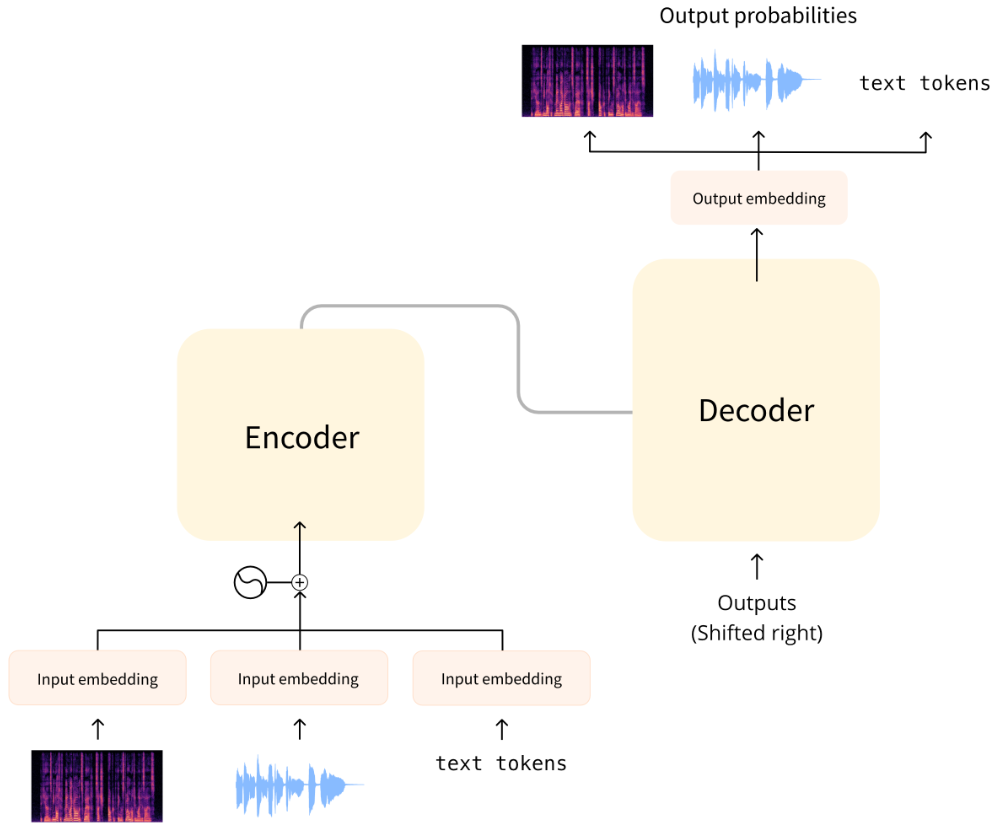}
    \caption{A basic overview of audio transformer encoders/decoders
    \cite{audio_transformers}.}
    \label{fig:audio_tf}
\end{figure}
\paragraph{Voice Based Control}
Voice controlled user interfaces can expand beyond the scope of language processing or conversational interactions to using vocal sounds for control. Voice based cursor control systems are one way to control activities on an interactive surface. By measuring use of vowel sounds, a user can move a cursor in an application without the need for a pointing device or gesture based commands \cite{10.1145/1168987.1169021}. These systems also provide added benefits for users with disabilities, who may have limited physical movements that make it difficult to use traditional interactive surfaces \cite{10.1145/1168987.1169021}. Additionally, emotions can be sensed from user speech patterns using features derived from spectral analysis of audio with machine learning techniques \cite{brady2016multi}. This can be used to adapt an interactive space to better reflect the mood of the user, or reconfigure the space if the user expresses feelings of frustration. 

\subsubsection{Hardware Level Sensing}
\paragraph{Standalone Acoustic Sensing}
By leveraging acoustic sensors in existing devices and with dedicated hardware, ordinary surfaces can be made interactive. One such system is SurfaceLink, which uses vibration motors in mobile phones to produce a low frequency tone \cite{goel2014surfacelink}. This tone can be heard by microphones in other phones, mapping the positions of each phone on a table, allowing for actions such as photo sharing simply by placing two phones closer together. Ordinary glass panes and walls can be turned into touch surfaces with acoustic sensors. By placing multiple microphones or piezoelectric sensors on a surface, the sound of a finger touching a surface can be triangulated, allowing for virtual user interfaces to be projected onto a surface without the need for large capacitive sensing arrays \cite{1037150}. 
\paragraph{Acoustic Augmented Sensing}
Acoustic sensing can be combined with other sensing modalities to enhance the user experience. Traditional capacitive touch sensing can capture the location and contact area of user gestures, but are unable to differentiate between a knock and a punch. By augmenting capacitive touch with microphones, the sound of the user touching the surface can be captured, providing additional context to differentiate between hand movements \cite{lopes2011augmenting}. More advanced systems expand this further: with machine learning classifiers, it is possible to identify when an input is made by a finger, a stylus, or even a ping pong ball \cite{harrison2011tapsense}. Acoustic aided sensing does not need to be directly incorporated into hardware designs; classifiers can run as part of software applications on existing devices such as phones and tablets, adding touch modalities beyond human inputs \cite{harrison2011tapsense}.
\begin{figure}
    \centering
    \includegraphics[width=1\linewidth]{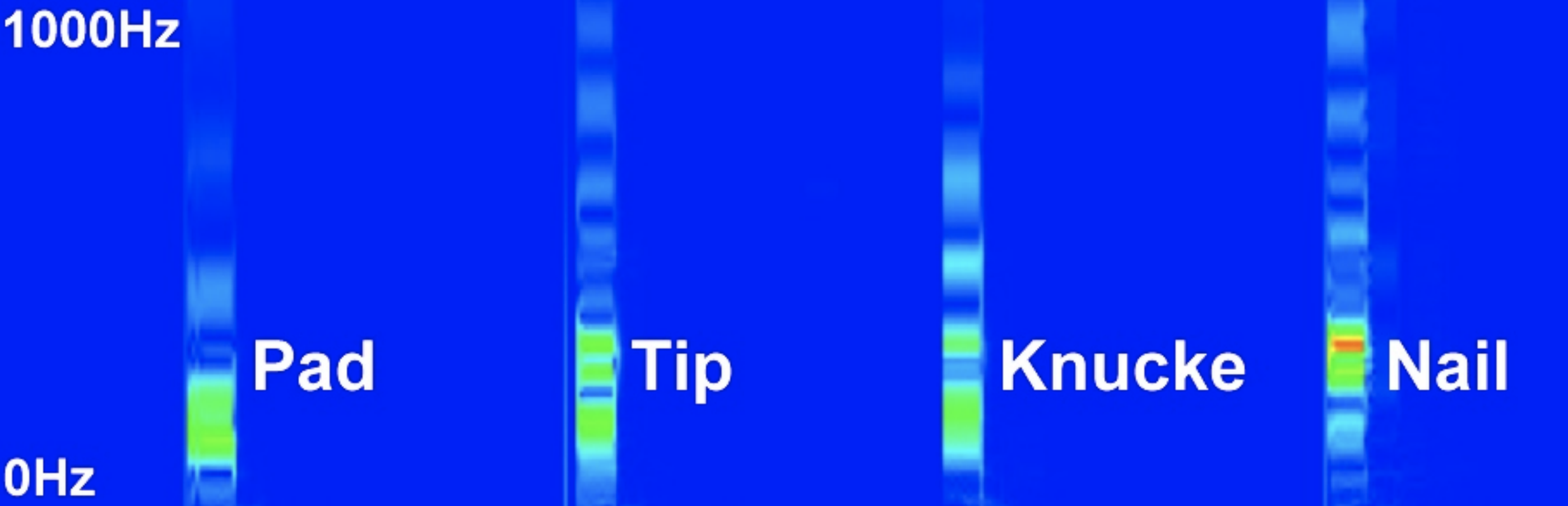}
    \caption{Spectrograms of audio waveforms created by different finger input types
    \cite{harrison2011tapsense}.}
    \label{fig:spectrum}
\end{figure}

\section{Present Challenges}
Several challenges exist in present implementations of interactive surfaces, including accuracy of sensing modalities, power consumption, and user security and privacy.

\subsection{Accuracy}
Sensing accuracy is crucial to the user experience, but challenges exist in making vision and acoustic sensing work for people of all backgrounds. Computer vision techniques depend heavily on the quality of the data that it trains on and the quality of its inputs (garbage in, garbage out). Camera resolution and frame acquisition rates can impact the fidelity of the application's results \cite{Wilson2005}. Real world environments are often noisy and inconsistent. Ambient lighting conditions will be different for each user and time of day. Conversations and ambient sounds from various sources decreasing sensing accuracy. Additionally, many audio detection models are tested in controlled environments with specific geometries and microphone placements, but in scenarios where sounds are reflected or absorbed by surfaces, audio detection models struggle to generalize. Every user is different as well, which can impact algorithm behavior. For example, computer vision algorithms have struggled to perform skin detection in complex environments or those that have similar colors\cite{HandGesture2020}. Improvements will also be needed to account for individuals who may not be represented well in the majority of training datasets.

\subsection{Power}
Power is a challenge for interactive surfaces, as computationally intensive vision and localization algorithms require powerful hardware to run. Additionally, wearable elements of interactive surfaces need to deliver accurate data, while having battery lives long enough to maintain a consistent user experience. Having to be connected to a static power source or a large portable source defeats the purpose of modern interactive surface philosophy. The conflict between power, performance, and area cause physical design challenges and redefinitions in the device's intended use. As such, modern design trends in hardware have focused on power efficiency to make the most of the limited constraints.

\subsection{Security \& Privacy}
Interactive surfaces are filled with sensors that record potentially sensitive user activities, especially those that use vision and movement based sensing methods, making security and privacy a top concern. Camera usage often makes users skeptical of the application's usage and can be an intrusion into private spaces. The same issues come with audio sensing, especially for interactive chat based NLP systems, where recordings of users may need to be offloaded to remote servers for processing. Users feel uncomfortable that they are being recorded, even in situations where recorded speech is not directly used. Interactive systems must give the user control over what is collected and be informed about how it will be used. New sensing modalities such as mmWave and vibration sensing aim to address this by providing accurate data while protecting user privacy.

\section{Emerging Technologies}
While most contemporary interactive surfaces make use of infrared, capacitive, acoustic and vision based sensing, new technologies such as mmWave and vibration sensing have potential to greatly improve the user experience. Using high frequency radio waves, mmWave sensing measures signal returns to determine the distance, velocity, and motion of objects with fine resolution. Vibration sensing detects changes in motion or mechanical oscillations by converting tiny movements into measurable electrical signals that reflect frequency, amplitude, and pattern.
\subsection{mmWave}
\subsubsection{Background}
Emerging mmWave sensing techniques, such as mmWave radar for multi-object movement tracking, are able to provide detailed information on the movement and position of objects within an interactive space, offering improved accuracy over vision based systems. Compared to traditional vision based tracking, mmWave presents several major improvements. As mmWave relies on radar, no cameras are needed, and mmWave systems are able to perform effectively in all lighting conditions, even if objects are partially obstructed by cloth or thin plastic \cite{zhang2023survey}. Additionally, mmWave does not need to store and process images of users, improving privacy and security \cite{zhang2023survey}. While past mmWave systems were bulky and required large antennas for sensing, newer systems are able to combine the antenna, RF front end, and signal processing into a single integrated circuit \cite{Lien_Gillian_2020}, making it much easier for mmWave radar to be integrated into mobile devices and smart spaces.

\subsubsection{Applications}
\paragraph{Movement Tracking} One of the biggest advantages mmWave provides is its high tracking accuracy. mmWave radar solutions have been used for tracking gestures and movements with sub-millimeter level accuracy at frame rates as high as 10000 FPS \cite{lien2016soli}. Users can use gestures in midair, such as pinches to zoom in or out, or carefully move their finger to scrub in a video, with precision levels exceeding that of capacitive touch based interactive surfaces. To achieve the accuracy needed to track these precise actions, many mmWave systems make use of beam steering techniques to direct the radar signals at the user, and use both analog and digital signal processing to extract features which are used for machine learning classifiers \cite{lien2016soli}. Additionally, mmWave radar is also able to effectively track large scale movements and multiple objects. Such systems can be used to track the positions of multiple users inside a smart environment, using techniques such as Kalman filtering and point clouds to achieve accuracy exceeding 90\% \cite{pegoraro2021real}. 
\begin{figure}
    \centering
    \includegraphics[width=1\linewidth]{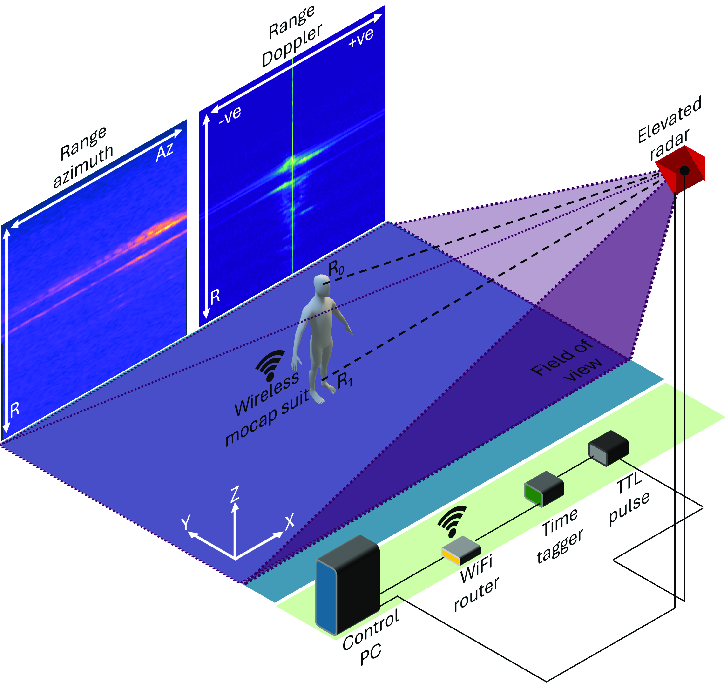}
    \caption{An example of a full body motion tracking system with mmWave radar
    \cite{10638541}.}
    \label{fig:mmwave}
\end{figure}

\paragraph{Orientation Detection} Using the Angle of Arrival of incoming mmWave signals, the orientation of objects can be accurately determined with geometric relations. These systems typically rely on multiple-input and multiple-output (MIMO) antenna arrays, with a transmitter located on an object for tracking, and a receiver built into a smart device or interactive space, which can reconstruct orientations with trigonometry based off multiple signals\cite{8240645}. This can be used to track the orientation of wearables on a user, or be combined with position data to precisely simulate real world objects in a virtual environment. Apart from sensing, mmWave is also used for high bandwidth communication, as it is able to transit data such as large video streams with low latency, making it a good fit for extended reality solutions. In this role, mmWave can serve a dual purpose, for orientation tracking as well as data transfer. Orientation data can be used to improve user tracking in an interactive space, or it can be used to improve mmWave data transfer rates via beamforming \cite{9994740}.

\paragraph{Mapping} While most high precision mapping systems use LiDAR to scan environments, mmWave can be used as a lower cost alternative, while offering reasonable mapping accuracy. Projects such as milliMap feed raw mmWave angle and range data into Generative Adversarial Networks, which can reconstruct a dense grid map for user in various applications \cite{lu2020see}. Mapping systems can be used by interactive spaces to obtain physical characteristics of users and any objects they may be carrying, improving interaction and assistance.

\subsection{Vibration}
\paragraph{Background} Vibration sensing functions similarly to audio sensing, but instead of processing sound waves, vibration sensing makes use of mechanical waves created when an object is touched or bent. These waves propagate throughout an object, and can be detected remotely with devices such as accelerometers, strain gauges, piezoelectric sensors, and geophones \cite{ma2025review}. Compared to traditional systems such as vision, capacitive touch, or audio sensors, vibration sensors have the advantage of not requiring direct line of sight or direct physical contact, while being less affected by noise.
\begin{figure}
    \centering
    \includegraphics[width=1\linewidth]{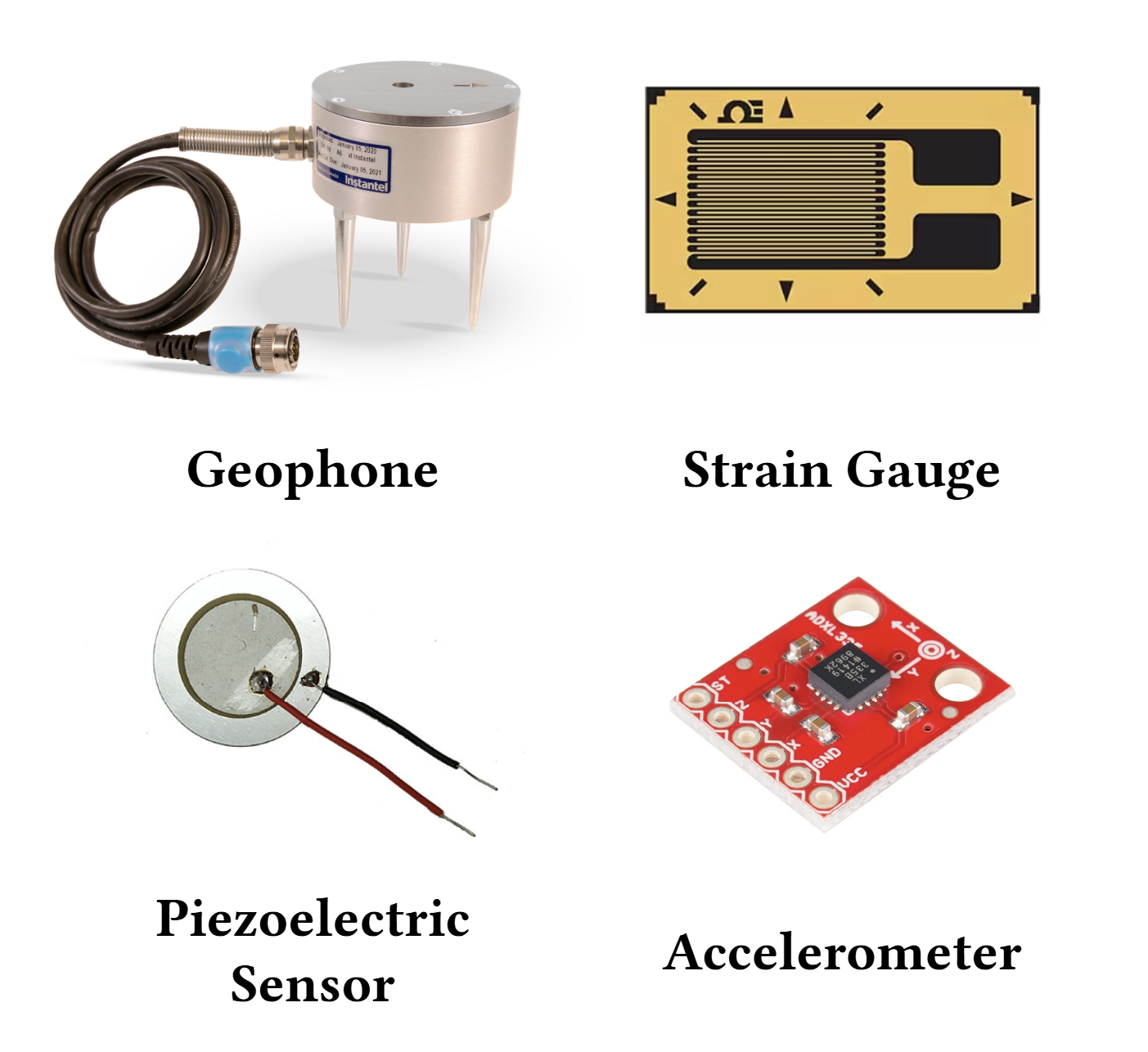}
    \caption{Common types of vibration sensors.}
    \label{fig:vibration}
\end{figure}
\subsubsection{Applications}
\paragraph{Touch Sensing}
 Smart vibration sensing systems built into interactive surfaces provide new user interfaces by localizing the position and strength of touches, allowing virtual tactile interfaces to be projected into the real world \cite{liu2017vibsense, zhang2025wevibe, wu2025humanbodyweightestimation}. Vibration sensing provides additional data beyond that of capacitive and audio sensing techniques. Capacitive sensors aren't able to capture force data, while noise makes it difficult for audio sensing to capture light touches. Instead, by using Voice Pickup Units to detect a type of vibration that travels along the surface of objects, known as Surface Acoustic Waves, it is possible to detect force sensitive events with up to 97\% accuracy, as shown in the SAWSense project \cite{10.1145/3544548.3580991}. Vibration sensors can also serve as a direct replacement for audio sensors in some scenarios, such as transforming ordinary surfaces into large scale touch sensors, but with added benefits. Audio sensing can localize taps and knocks on these surfaces, but can struggle with when used with different materials, and is unable to track gestures such as swipes. Systems such as SurfaceVibe use multiple geophones mounted to a surface and measure the Time Difference of Arrival (TDoA) to precisely locate touches on an interactive surface \cite{7944790}. Furthermore, by localizing multiple signal segments, it is possible to determine the trajectory of touch inputs and therefore detect when a user swipes across the surface \cite{7944790}.

\paragraph{Movement Tracking}
Using vibrations, it is possible to gain insight into individual movement characteristics, such as gaits, that other sensing modalities aren't able to effectively capture. FootprintID is one such system that utilizes unique structural vibrations induced by people walking through buildings to identify individuals. By using Refined Transductive and Iterative Transductive Support Vector Machines (RTSVMs and ITSVMs) with vibration data collected from geophones, the system is able to identify individuals with up to 96\% accuracy, even when walking speed and location changes \cite{10.1145/3130954}. While indoor localization with vibration sensors is challenging due to the high velocities of mechanical waves, instead of directly using vibrations to detect user positions, it is possible to track vibrations induced by user actions and estimate a movement trajectory for the user \cite{kashimoto2016floor}. For example, if a person sits down on a chair or opens a door in an interactive space, the vibrations produced by these actions can be detected with a single piezoelectric sensor, and be fed into a classifier to reminder the action that occurred \cite{kashimoto2016floor}. As the chair and door are in known positions, it is possible to estimate the location of the user, without the need for cameras or IR systems, improving privacy.

\paragraph{Wearables}
Wearable devices for vibration monitoring offer innovative solutions to sensing challenges. In contrast other approaches in which sensors are mounted on a surface to make it touch sensitive, one idea is to instead have the user wear the sensors. Using piezoelectric polymer films, it is possible to detect vibrations transmitted through skin when a user touches an object \cite{tanaka2015wearable}. As vibrations travel through the entire hand when touches occur, the polymer sensing film only needs to be placed on a single finger, improving comfort for the user \cite{tanaka2015wearable}. A major challenge with wearables is battery life, which has inspired efforts to passively harvest energy to power devices. One technique is Kinetic Energy Harvesting (KEH), which uses piezoelectric elements to generate power from vibrations. KEH systems can serve a dual purpose for both energy generation and vibration sensing. By performing feature extraction on the power signal produced by a KEH and using feature data to classify activities, it is possible to determine if users perform actions such as running, walking, jumping, or climbing, all with a passively powered device that never needs charging \cite{khalifa2017harke}. Passively powered activity trackers ease the frustration of having to recharge wearable accessories, and can feed motion activity data to an interactive space without the need for cameras.

\section{Conclusion}
Interactive surface technologies have evolved from early optical and infrared techniques to a wide range of sensing approaches, including capacitive, acoustic, mmWave, and vibration-based methods. Each modality offers distinct strengths whether in precision, scalability, robustness, or ease of integration while also presenting trade-offs that continue to shape how users interact with digital content. As advances in materials, embedded hardware, and signal processing accelerate, multimodal sensing is emerging as a promising direction for achieving more reliable, expressive, and context aware surface interactions.

Despite these developments, several challenges remain open. Differentiating multiple simultaneous users, enabling robust noncontact interaction, adapting to diverse materials and environments, and balancing privacy with sensing fidelity all require further research. Looking ahead, progress in lightweight, scalable, and seamlessly integrated sensing technologies will be key to enabling truly ubiquitous interactive environments, where everyday surfaces become active participants in human computer interaction.

\clearpage
\bibliographystyle{ACM-Reference-Format}
\bibliography{references}
\appendix
\section{Interactive Surfaces Sensing Summary}

\begin{table*}[t]
\centering
\begin{tabular}{p{3cm} p{5.5cm} p{5.5cm}}
\hline
\textbf{Modality} & \textbf{Strengths} & \textbf{Weaknesses} \\
\hline

Infrared (FTIR) &
Low cost and scalable for large surfaces; supports multi touch; simple optical setup &
Requires pressure for reliable activation; bulky form factor; sensitive to ambient light and surface contamination; limited precision for fine inputs \\
\hline
Diffuse Illumination (Rear and Front DI) &
Works on many surface materials; enables large, low cost interactive surfaces; supports both touch and object recognition; no need for light guiding substrates &
Sensitive to ambient illumination; shadow and reflectance artifacts cause false touches; camera resolution limits accuracy; environmental noise impacts sensing \\
\hline
Capacitive Touch &
High precision and low latency; not affected by lighting; thin form factor; mature and inexpensive; strong multi touch support &
Can't measure force; sensitive to moisture; performance varies across materials; flexible implementations introduce drift and durability issues \\
\hline
Computer Vision &
Enables complex interactions including midair gestures; supports hand, body, and object tracking; scalable to room sized environments &
Accuracy depends on lighting, camera quality, and noise; privacy concerns with images/video; computationally expensive; sensitive to occlusion \\
\hline
Acoustic (Voice Interfaces) &
Supports hands free interaction; leverages NLP and modern ASR; beneficial for accessibility; can detect user intent and emotion &
Degrades in noisy environments; privacy concerns with continuous audio capture; model accuracy varies across accents and speaking styles \\
\hline
Acoustic (Hardware Sensing) &
Enables touch detection on arbitrary surfaces; supports object classification based on contact sounds; minimal hardware required &
Material dependent performance; ambient noise disrupts accuracy; limited precision for fine gestures \\
\hline
mmWave &
High tracking accuracy; robust in all lighting; works through some obstructions; preserves privacy better than cameras; supports gesture, movement, and orientation sensing &
Complex signal processing; environmental multipath effects can complicate tracking; hardware still maturing for consumer products \\
\hline
Vibration &
Works through opaque materials and without line of sight; detects force and subtle touch dynamics; low power; supports large touch sensing areas and wearable interaction &
Strongly material dependent; mechanical wave propagation complicates localization; limited spatial resolution without multiple sensors; wearable implementations must manage comfort and power \\

\hline
\end{tabular}
\caption{Summary of sensing modalities in interactive surface technologies, with key strengths and weaknesses.}
\end{table*}

\end{document}